\pdfoutput=1

\newcommand{\dx}{\text{d}x}
\newcommand{\dy}{\text{d}y}
\newcommand{\ner}{\mathbf{r}}
\newcommand{\nrc}{\ner_c}
\newcommand{\Ndel}{G}
\newcommand{\dd}{\text{d}}

\documentclass[11pt,letter]{article}

\usepackage{float}
\usepackage[numbers]{natbib}

\usepackage[colorlinks=true,allcolors=blue]{hyperref}
\usepackage{latexsym}
\usepackage{amsmath}
\usepackage{amsfonts}
\usepackage{amssymb}
\usepackage{amsthm}
\usepackage{graphicx}
\usepackage{empheq}
\usepackage{caption}
\usepackage{enumerate}
\usepackage{color}
\usepackage{subfig}
\usepackage{mathrsfs}

\begin{document}

\title{Soliton dynamics in finite nonlocal media with cylindrical symmetry}

\author{Emmanuel Garza$^1$, Servando Lopez-Aguayo$^2\footnote{E-mail: \texttt{servando@itesm.mx}}$,
  and Julio C. Guti\'{e}rrez-Vega$^2$
  \\ \\
  \small{$^1$Computing and Mathematical Sciences, California Institute of
  Technology, Pasadena, California 91106, USA} \\
  \small{$^2$Photonics and
  Mathematical Optics Group, Tecnologico de Monterrey, Monterrey 64849, Mexico}
}

\date{\today}

\maketitle

\begin{abstract} 
  The effect of finite boundaries in the propagation of spatial nonlocal
  solitons in media with cylindrical symmetry is analyzed. Using Ehrenfest's
  theorem together with the Green's function of the nonlinear refractive index
  equation, we derive an analytical expression for the force exerted on the
  soliton by the boundaries, verifying its validity by full numerical
  propagation. We show that the dynamics of the soliton are determined not only
  by the degree of nonlocality, but also by the boundary conditions for the
  refractive index. In particular, we report that a supercritical pitchfork
  bifurcation appears when the boundary condition exceed a certain threshold
  value.
\end{abstract}

\section{Introduction}
Nonlinear self-trapped optical beams, also known as spatial optical solitons,
have been extensively studied over the past years in several media for their
potential use in all-optical communication
networks~\cite{Kivshar2003xv,Chen2012}.  Of special interest is the study of
nonlocal solitons, which are present when the nonlinear response at a particular
point of the medium depends not only on the optical field localized at that
single point, instead, the response is a function of the intensity around the
neighboring region. Some examples of media with nonlocal processes are the
diffusion of charge carriers~\cite{Skupin2006}, thermal
media~\cite{Kartashov2007b}, and liquid crystals~\cite{Peccianti2007}.
Nonlocality allows several new phenomena, that are not possible to observe in
pure Kerr local media. For example, nonlocality can form new bound states and
stable families of solitons such as Hermite solitons~\cite{Ma2011}, Laguerre
solitons~\cite{Buccoliero2007}, azimuthons~\cite{Lopez-Aguayo2006a} and
ellipticons~\cite{Lopez-Aguayo2007}, among other self-trapped nonlinear optical
beams. Additionally, the presence of finite boundaries in nonlocal media can
induce transformations between solitons of different symmetries and exert
repulsive forces ~\cite{buccoliero2009boundary,louis2013optical}. In fact, there
is the possibility that far-away asymmetric boundary forces can exert control
into the soliton dynamics as it was shown in~\cite{Alfassi2007} for the case of
propagation in highly nonlocal nonlinear media.  Similarly, a power-dependent
nonlinear repulsion at the boundary was reported in~\cite{Alberucci2007a} and
for the case of thermal optical nonlinearity, the range of action of the
boundary is virtually infinite~\cite{Rotschild2006b}. When this range of action
is much larger than the width of the soliton, equivalent particle theory can be
used to describe the soliton dynamics~\cite{particle_theory}. In particular,
using the method of images, and assuming that the soliton waist is much smaller
than the radius of the boundary, Shou and co-workers~\cite{Shou2009a} obtained
an analytical approximation for the boundary force exerted on a soliton for the
case of thermal self-focusing nonlinearity in lead glass with circular
boundaries, which corresponds to the highly nonlocal limit. In this report, we
extend the analysis of the circular cylinder to include the case of an arbitrary
degree of nonlocality, as well as general boundary conditions for the
nonlinear-nonlocal refractive index. We report an analytical expression for the
boundary force that shows good agreement with numerical results, revealing that
soliton dynamics for a general degree of nonlocality, are significantly
different from the purely nonlocal limit case.

\section{Physical model}
We start our analysis by considering the nonlinear propagation of a monochromatic complex
field envelope $U(X,Y,Z)$ in a nonlocal medium described by the following
nonlinear Schr\"odinger equation
\begin{align} \label{eqn:paraxial0}
2ik \frac{\partial U}{\partial Z} +\bigg( \frac{\partial^2}{\partial X^2}+ \frac{\partial^2}{\partial Y^2}\bigg) U + 2 k^2 \frac{\Delta n(|U|^2)}{n_0} U = 0,
\end{align}
where $Z$ is the propagation coordinate, $X,Y$ are the transverse coordinates,
$n_0$ is the linear refractive index of the medium, $k=\omega n_0/c$ is the wave
number, and $\Delta n$ represents the nonlinear change in the refractive index
that depends upon the optical field intensity distribution. Here we study the
case where the nonlocal nonlinear response of the medium obeys the 2D screened
Poisson equation~\cite{Alberucci2007}
\begin{align} \label{eqn:poisson0}
b \bigg( \frac{\partial^2}{\partial X^2}+ \frac{\partial^2}{\partial Y^2}\bigg) \Delta n-a \Delta n  + | U |^2 =0 ,
\end{align}
where $a$ and $b$ are parameters that depends on the particular properties of
the medium. The model represented by~\eqref{eqn:poisson0} has been previously
used to describe partially ionized plasmas~\cite{Yakimenko2005}, plasma heating
on the propagation of electromagnetic waves~\cite{Skupin2006} and thermal
nonlinear process in the regime of strong absorption
\cite{ghofraniha2007shocks}. In particular, a physical counterpart of the model
given by ~\eqref{eqn:poisson0} is used to study nematic liquid crystals in a
planar cell that accounts for the observability of accessible
solitons~\cite{Alberucci2007,conti2003route,assanto2003nematicons}. The ratio
$a/b$ can be related to the degree of nonlocality; the case $a/b\rightarrow0$
corresponds to highly nonlocal media, while $a/b \rightarrow\infty$ models pure
Kerr media.  We restrict ourselves to the case on which the medium has a finite
cross section given by a circular boundary of radius $R_b$. Without loss of
generality, we rescale the spatial coordinates by the relations $x = X /R_b$,
$y = Y/R_b$, $z = Z/(kR_b^2)$, and introduce the normalized versions of the
nonlinear response, field envelope and nonlocal parameter given by
$N = k^2 R_b^2 \Delta n /n_0$, $\Psi = k R_b^2 U/\sqrt{b n_0}$ and
$\rho = \sqrt{a/b}R_b$, respectively. Additionally, given the circular symmetry
of the boundaries, we use cylindrical coordinates $(r,\theta,z)$ with
$r=\sqrt{x^2+y^2}$, $\theta = \arctan{(y/x)}$ and the position vector at a fixed
$z$ is denoted by $\mathbf{r} = x \hat{i} + y \hat{j}$.  Under these
assumptions, the governing equations for the beam
propagation~\eqref{eqn:paraxial0}~and~\eqref{eqn:poisson0} are now given by
\begin{subequations}
\begin{align} \label{eqn:paraxial1}
  \begin{cases}
    \displaystyle i \frac{\partial \Psi}{\partial z} + \frac{1}{2} \nabla_\perp^2 \Psi + 
    N(|\Psi|^2) \Psi = 0, & r\le 1 \; \text{and} \; z>0, \\
    \Psi(\mathbf{r},z) = \Psi_0(\mathbf{r}), & z = 0,
  \end{cases}
\end{align}
\begin{align} \label{eqn:poisson}
  \begin{cases}
    \rho^2 N - \nabla_\perp^2 N = | \Psi |^2,  & r < 1 \; \text{and} \; z\ge 0,\\
    N(\mathbf{r},z) = N_b(\theta), & r=1  \; \text{and} \; z \ge 0,
  \end{cases}
\end{align}
\end{subequations}
where $\nabla_\perp^2 = \partial^2/\partial x^2 + \partial^2 / \partial y^2$ stands for the
transverse Laplacian.
\section{Results}
\subsection{Analytical results}
We start our analysis by finding fundamental solitons centered at the origin by
using the ansatz $\widetilde \Psi(\ner,z) = e^{i\lambda z} \psi(\ner)$, where
$\lambda$ is the soliton's propagation constant and $\psi(\ner)$ is a real
function. Thus, \eqref{eqn:paraxial1} becomes
\begin{align} \label{eqn:nlseind}
\frac{1}{2} \nabla_\perp^2 \psi + N(|\psi|^2) \psi - \lambda \psi = 0,
\end{align}
while in \eqref{eqn:poisson} $\Psi$ is just replaced by $\psi$. Then, different
soliton trajectories can be analyzed by adding a tilt to
$\widetilde \Psi$. The position of the soliton at a fixed propagation
distance $z$, is characterized by the intensity centroid
$\nrc = x_c \hat{i} +y_c \hat{j}$ using
\begin{equation}
\nrc(z) = \frac{1}{P} \int \int |\Psi(\ner,z)|^2 \; \ner \; \dx \dy,
\end{equation}
where the power of the soliton is 
\begin{equation}
P=\int \int |\Psi|^2 \dx \dy.
\end{equation}
By introducing a finite medium where the refractive index is a solution to
\eqref{eqn:poisson}, the launched soliton experiences a force whenever the beam
center is not at the origin due to the asymmetry in the refractive index $N$
generated in order to meet the boundary conditions. The force exerted to the
beam centroid by the boundaries can be described by Ehrenfest's
theorem~\cite{Alberucci2007}
\begin{align} \label{eqn:forcenum}
\frac{\text d^2 \nrc }{\text d z^2} = 
\frac{1}{P}\int \int |\Psi(\ner,z)|^2 \nabla N(\ner, z) \dx \dy.
\end{align}
For the highly nonlocal limit, and assuming that the soliton waist is much
smaller than the radius of the waveguide, an analytical approximation to the
boundary force was obtained in~\cite{Shou2009a} using the method of images. This
approximation is based on the assumption that the entire power of the beam is
localized at the intensity centroid, i.e.
$|\Psi(\ner,z)|^2 \approx P \delta(\ner-\nrc(z)) $, where $\delta(\ner-\nrc)$ is
the Dirac delta function. Under this assumption, the induced refractive index
$N$ is then given by the Green's function of~\eqref{eqn:poisson} (for a fixed
$z$). In the case $\rho = 0$ (Poisson's equation), the method of images can be
used to find the Green's function, but for arbitrary $\rho$ we need to consider
a different procedure.  Assuming that the soliton can be model as
$|\Psi|^2 = P \delta(\ner-\nrc)$, the induced refractive index can be found by
solving the following boundary value problem
\begin{align} \label{eqn:poisson_del}
  \begin{cases}
     \rho^2 \Ndel(\ner,\nrc)-\nabla_\perp^2 \Ndel(\ner,\nrc)  =  
     P \delta(\ner-\nrc), & \text{for } r<1, \\
    \Ndel(\ner,\nrc) = N_b(\theta), & \text{for } r= 1,
  \end{cases}
\end{align}
which we solve by using the method of separation of variables alongside standard
techniques for finding Green's
functions~\cite{duffy2001green,Perez-Arancibia2010}. We proceed to split $\Ndel$
into three components
$\Ndel(\ner,\nrc) = \Ndel_1(\ner,\nrc) + \Ndel_2(\ner,\nrc) + \Ndel_H(\ner)$,
where $\Ndel_1(\ner,\nrc)$ is the Green's function of the screened Poisson
equation subject to the solution being finite at infinity, $\Ndel_2(\ner,\nrc)$
is such that $\Ndel_1(\ner,\nrc)+\Ndel_2(\ner,\nrc)$
satisfies~\eqref{eqn:poisson_del} with zero boundary conditions, and
$\Ndel_H(\ner)$ is a homogeneous solution that satisfies the boundary condition.
Each term is given explicitly by
\begin{align}
  \begin{cases}
    \displaystyle \Ndel_1(\ner,\nrc) = \frac{P}{2\pi} K_0(|\ner-\nrc|), \\
    \displaystyle \Ndel_2(\ner,\nrc) = - \frac{P}{\pi \rho~r_c } \sum_{n=-\infty}^\infty
    \frac{K_n(\rho) I_n(\rho r) I_n(\rho r_c)}{I_n(\rho) F_n(\rho r_c)} 
    \cos{[n(\theta-\theta_c)]}, \\
    \displaystyle \Ndel_H(\ner) = \sum_{n=-\infty}^\infty A_n 
    \frac{I_n(\rho r)}{I_n(\rho)} e^{i n \theta},
  \end{cases}
\end{align}
where $I_{n}$ and $K_{n}$ stand for the modified Bessel functions of the first
and second kind respectively, $A_n$ are the Fourier coefficients
of $N_b(\theta)$, that is $N_b(\theta) = \sum A_n e^{i n \theta}$, and
\begin{align}
  F_n(\rho r_c) = I_n(\rho r_c) [K_{n+1}(\rho r_c) + K_{n-1}(\rho r_c) ] + 
                    K_n(\rho r_c) [I_{n+1}(\rho r_c) + I_{n-1}(\rho r_c) ].
\end{align}
The term $\Ndel_1$ is symmetric with respect to the beam center and thus
provides the self-focusing effect that sustains the soliton~\cite{Shou2009a}. On
the other hand, the terms $\Ndel_2$ and $\Ndel_H$ produce a steering force on
the beam. Using~\eqref{eqn:forcenum} and the approximation
$|\Psi|^2 = P \delta(\ner-\nrc)$ we find that
\begin{subequations}\label{eqn:main_result}
  \begin{equation}\label{eqn:main_result1}
    \displaystyle \frac{\dd^2 \nrc}{\dd z^2} = \mathbf{F}_2(r_c) + 
    \mathbf{F}_H(r_c,\theta_c), 
  \end{equation}
  \begin{equation}\label{eqn:main_result2}
    \displaystyle \mathbf{F}_2(r_c) = -\frac{P}{2\pi r_c} \sum_{n=-\infty}^\infty
      \frac{K_n(\rho) I_n(\rho r_c) 
          [I_{n+1}(\rho r_c) + I_{n-1}(\rho r_c)] }{I_n(\rho) F_n(\rho r_c)}  
        \; \mathbf{\hat r},
  \end{equation}
  \begin{align}\label{eqn:main_result3}
    \displaystyle \mathbf{F}_H(r_c,\theta_c) = \sum_{n=-\infty}^\infty 
        \frac{A_n}{I_n(\rho)} e^{i n \theta_c} \left\{  \frac{\rho}{2} [I_{n+1}(\rho r_c) 
          I_{n-1}(\rho r_c)] \; \mathbf{\hat r} +  i\frac{n}{r_c} I_n(\rho r_c) 
          \boldsymbol{ \hat{ \theta } } \right\}.
  \end{align}
\end{subequations}
These equations constitute the main analytical result in this report and they
describe the effect of finite boundaries in the propagation of spatial nonlocal
solitons inside a circular cylinder.

\subsection{Numerical validation}
Next, we validate the quality of the analytical approximation by means of direct
numerical experimentation. In order to obtain fundamental soliton solutions, we
use a nonlinear least square method applied to~\eqref{eqn:nlseind} and we find
the optimal parameters of a centered Gaussian beam of a given waist $\omega_0$
for unknown amplitude $A$ and propagation constant $\lambda$. This way we can
control the width of the beam to satisfy $\omega_0 << 1$, which guarantees the
validity of~\eqref{eqn:main_result}. We find by means of numerical
experimentation that using the restriction $\omega_0 < 0.2$, the solitons
dynamics are described satisfactorily by our analytical results for at least of
$20$ diffraction lengths. The solution for the refractive index
in~\eqref{eqn:poisson} (for a fixed $z$) is found using a similar method to that
of~\cite{Lai2002a}, on which a Fast Fourier transform and a finite difference
scheme are used to discretize the polar and radial coordinates respectively. We
propagate the solitons by means of the split-step method~\cite{Weideman1986}.

\begin{figure}[!htb]
\centering{
\includegraphics[height=1.7in]{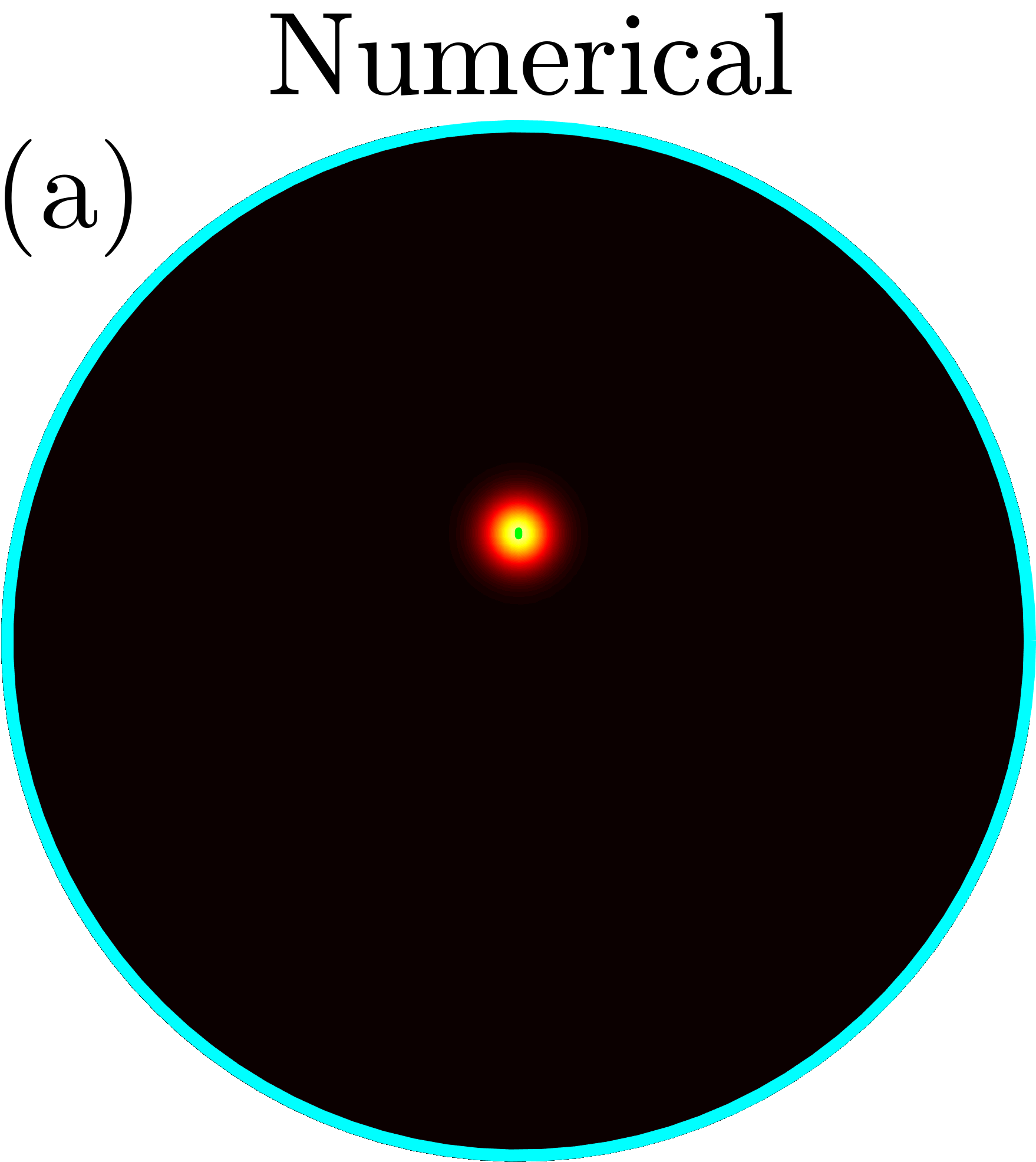} 
\includegraphics[height=1.7in]{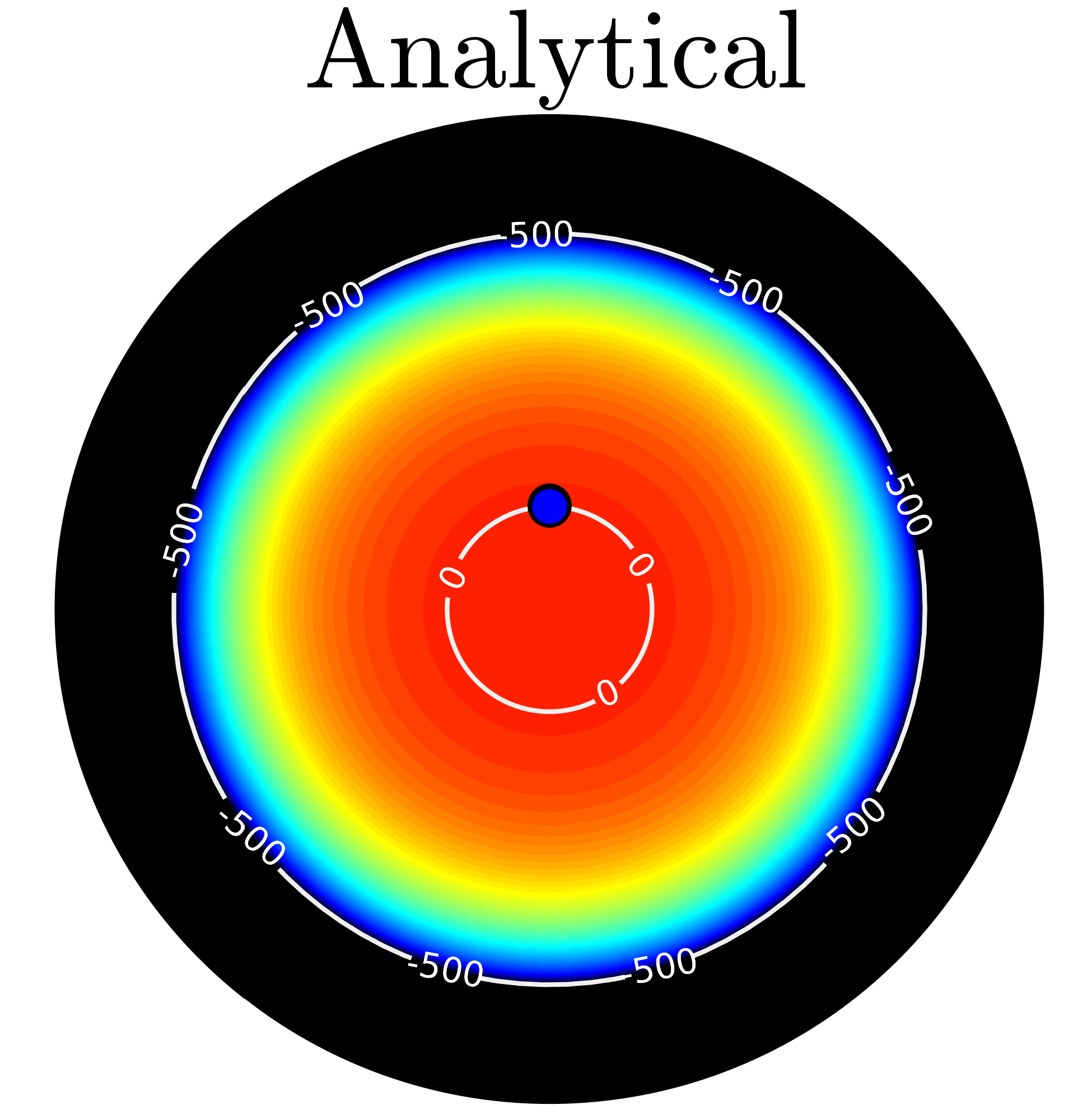} \\
\includegraphics[height=1.5in]{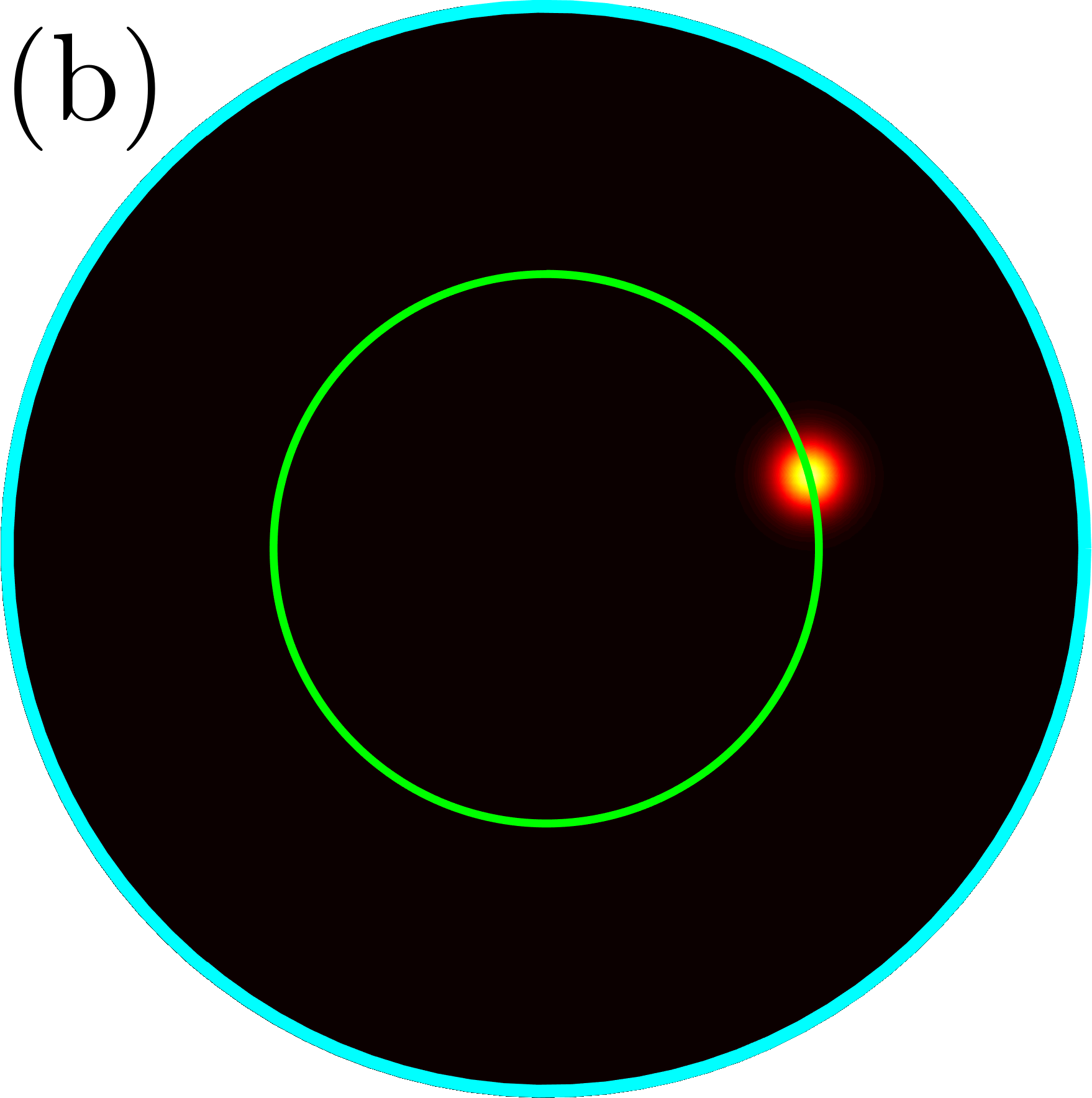} 
\includegraphics[height=1.5in]{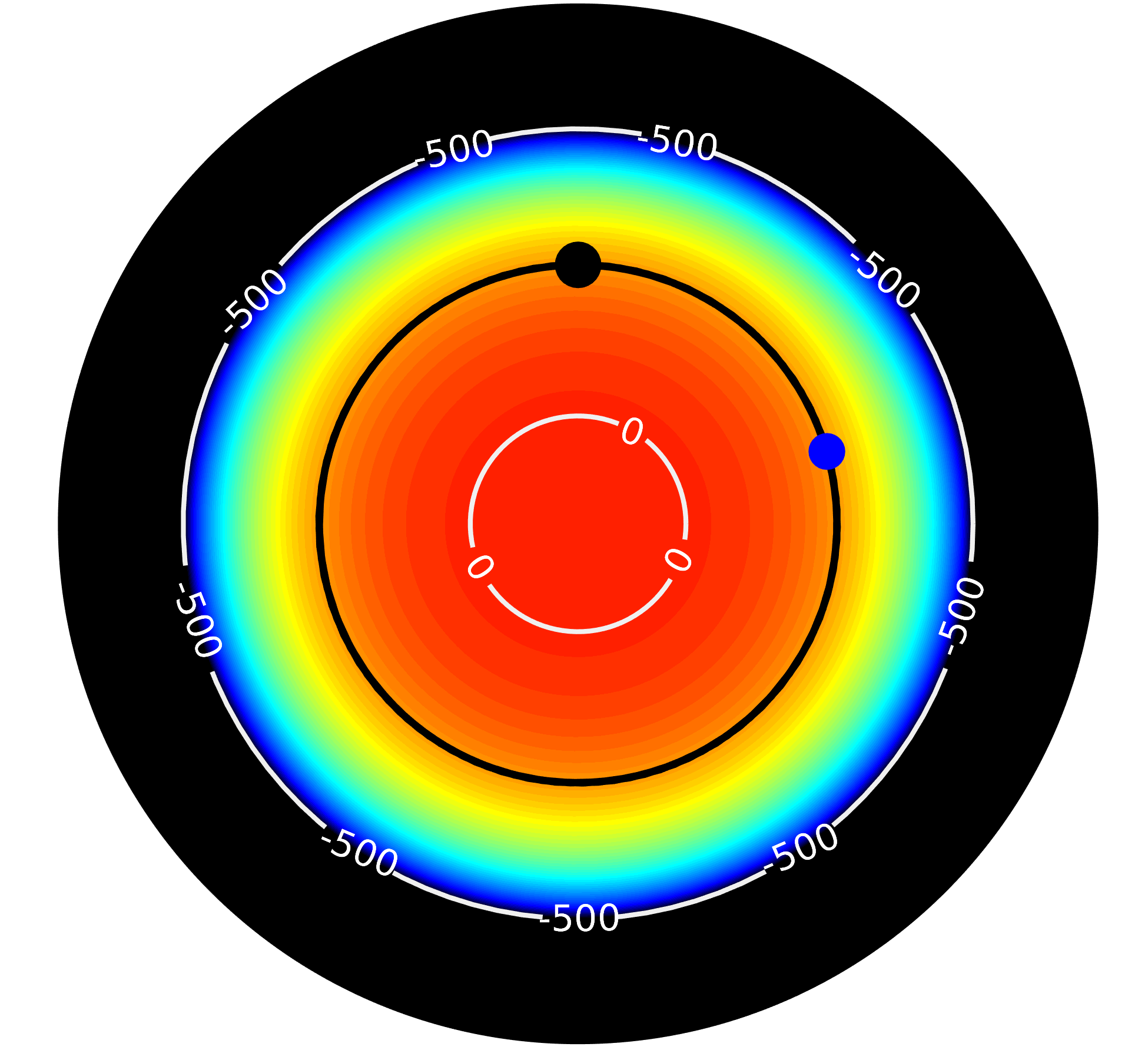} \\
\includegraphics[height=1.5in]{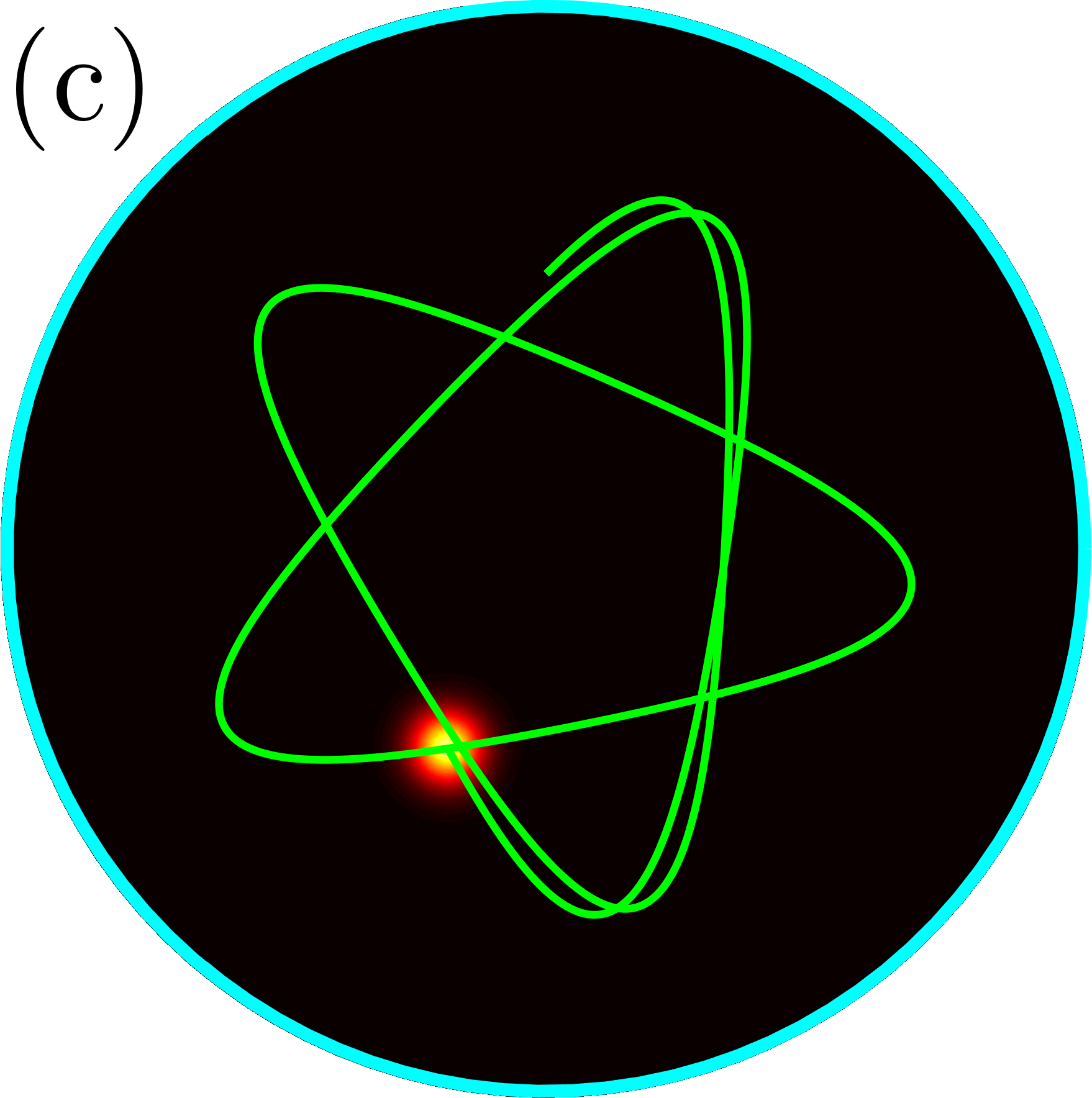} 
\includegraphics[height=1.5in]{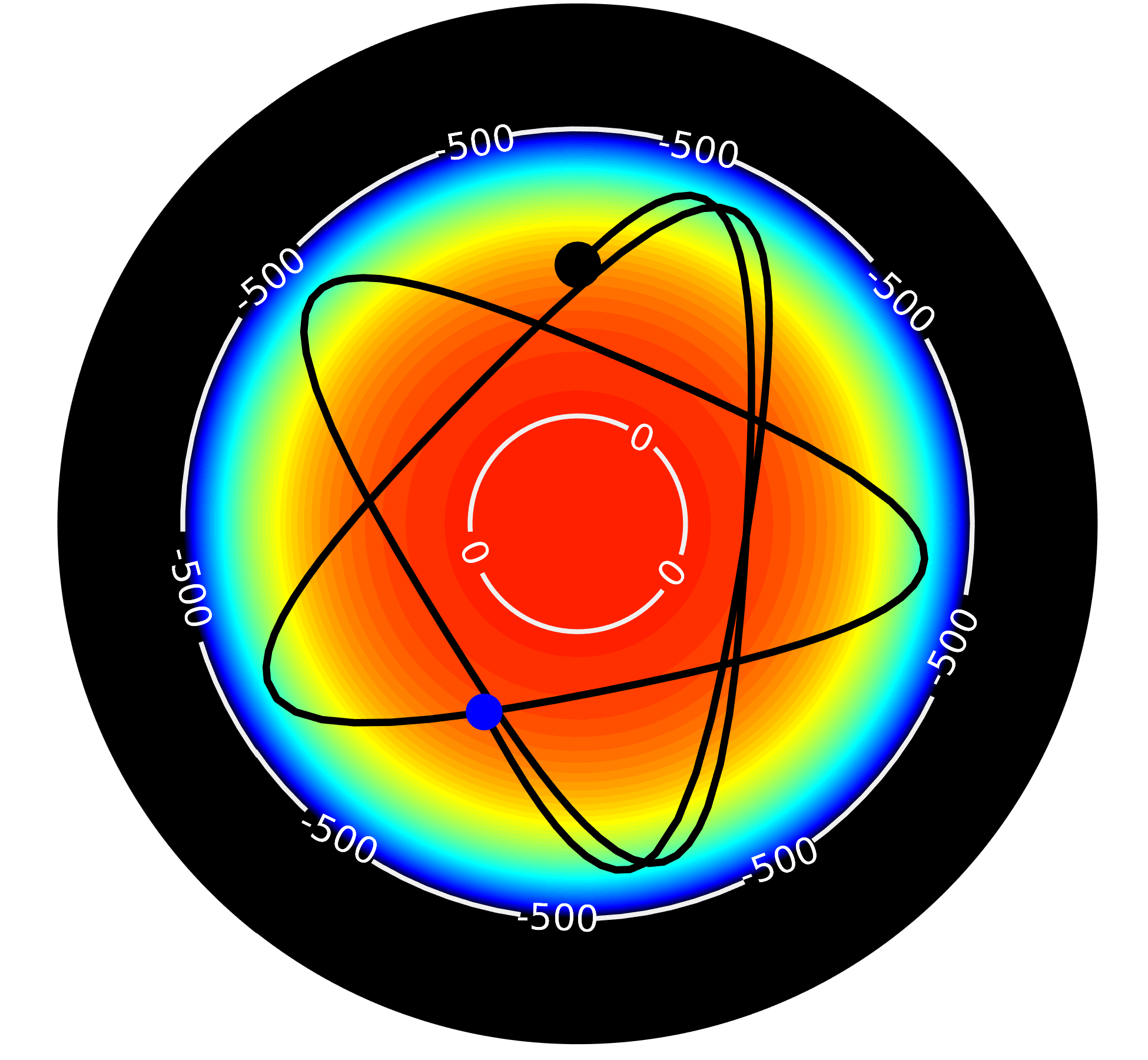} \\
\includegraphics[height=1.95in]{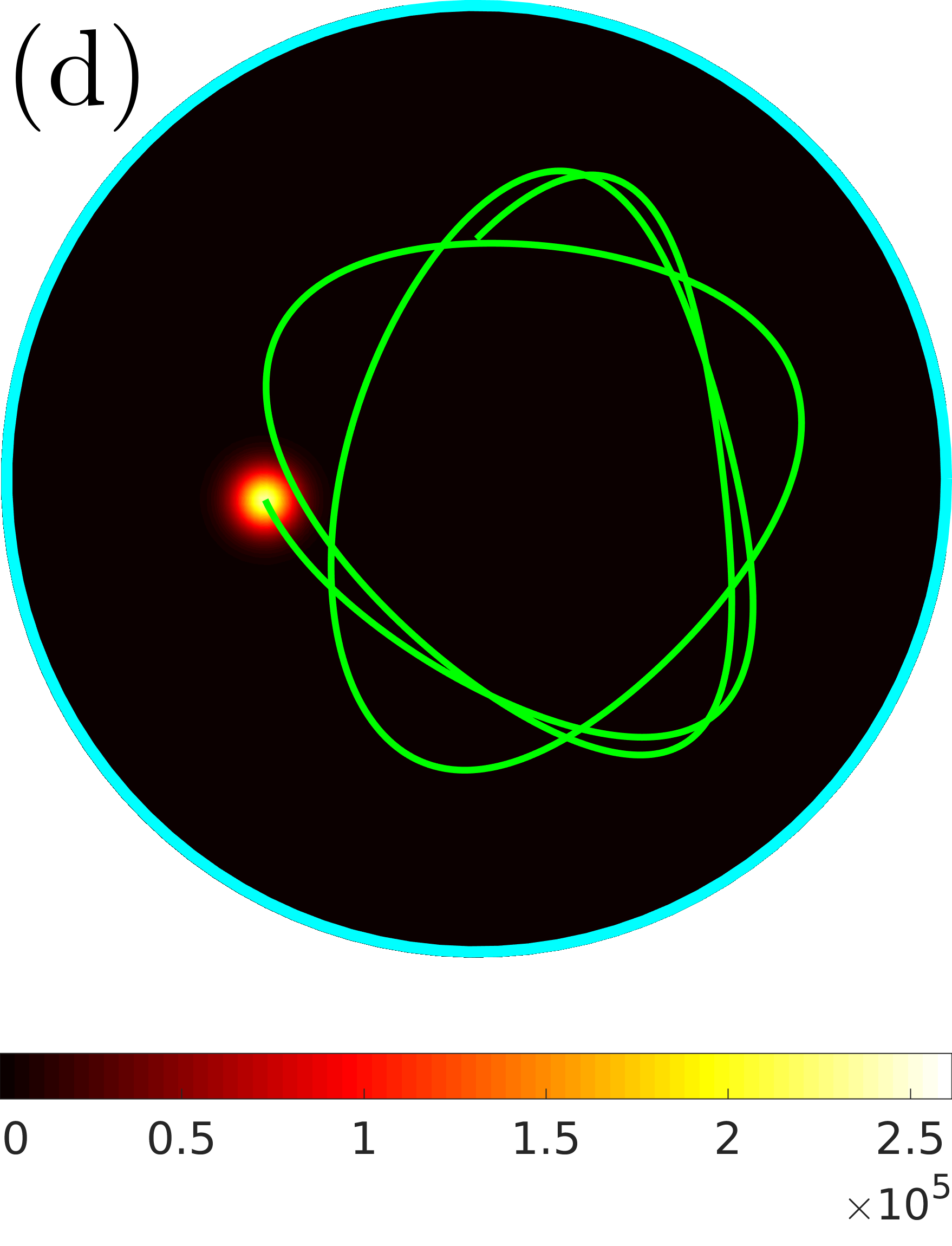} 
\includegraphics[height=1.95in]{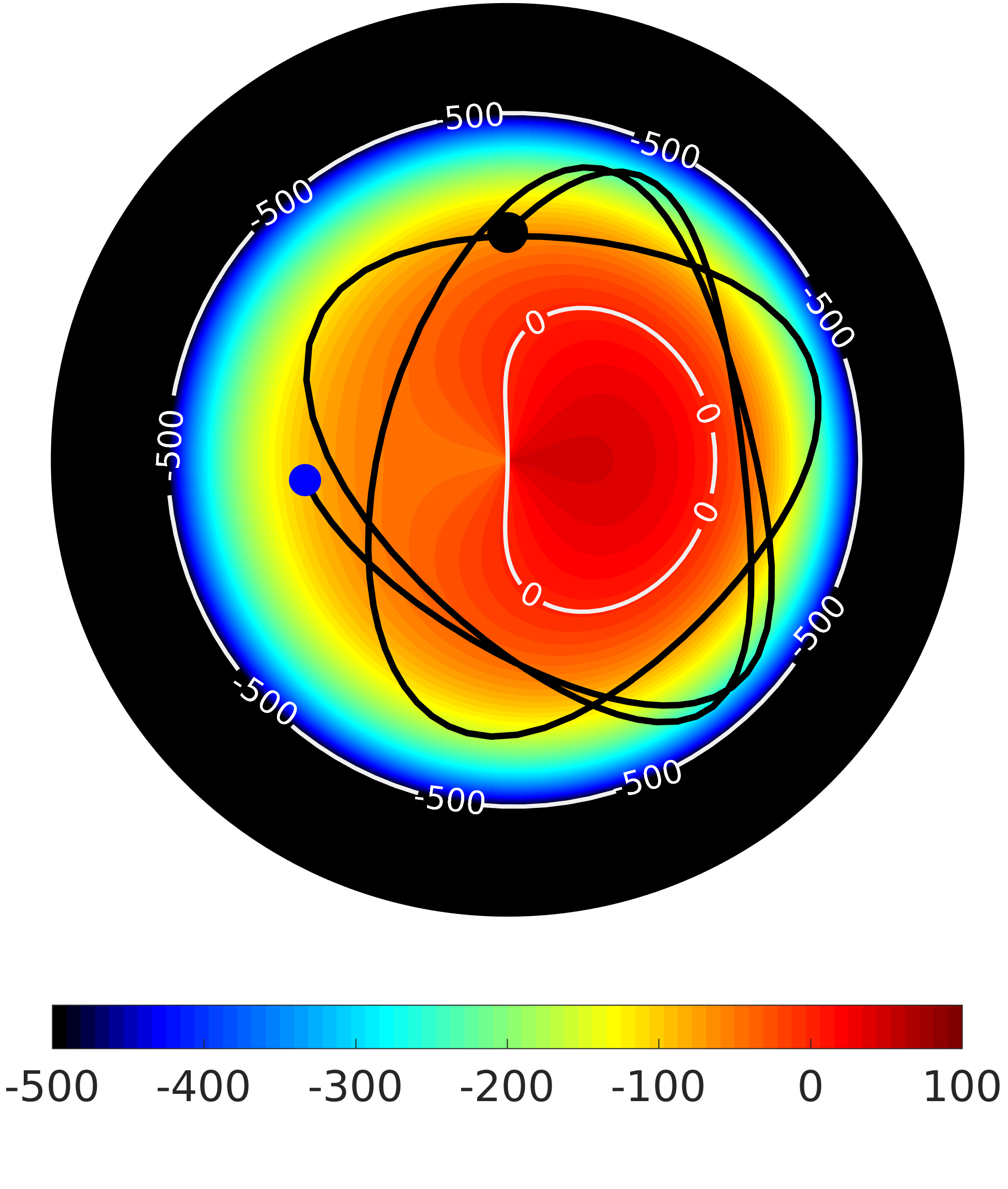} 
}
\caption{Soliton trajectories for a variety of launching conditions. The first
  column corresponds to the path given by the full numerical propagation (shown
  in green) and the colormap represents the beam intensity $|\Psi|^2$ at
  $z = 400\pi/ \lambda$. The second column presents the prediction by the
  analytical model~\eqref{eqn:main_result}, where the black curve is the
  predicted path of the soliton starting from the black marker and finishing on
  the blue one, and the colormap represents the radial force with contour lines
  of $0$ and $500$ also shown. For all cases $\rho=1$,
  $z\in[0,400 \pi/\lambda]$, $P=3000$. For (a)-(c) $N_b=900$ and in particular,
  (a) shows a static soliton centered at an equilibrium point $x_c=0.21$, (b)
  presents a circular orbit of radius $0.5$, (c) has a trajectory bounded by a
  minimum and maximum radii, and (d) has an added angular modulation at the
  boundary given by $N_b(\theta)=900+50 \cos(\theta)$.}
\label{fig:dynamics}
\end{figure}

In Fig.~\ref{fig:dynamics} we present several soliton trajectories predicted by
the analytical model (right column), as well as the corresponding full numerical
propagation (left column). First, we demonstrate that the centroid of the
soliton remains fixed as long as we launch it at an equilibrium point, as it is
shown in Fig.~\ref{fig:dynamics}~(a). On the other hand,
Fig.~\ref{fig:dynamics}~(b) shows a circular orbit obtained by launching the
soliton with an initial transverse momentum (tilt) balanced out with the
centripetal force generated by the boundary condition. The particular launching
conditions for this case are $r_c(z=0)=0.5$ and
$\Psi(\mathbf{r},0) = \Psi_0(\mathbf{r}-\ner_c(0),0) \exp{(ig_x x + i g_y y)}$
where we set $g_x=\sqrt{F_r r_c}=5.89$ and $g_y=0$. Similar soliton interactions
in a nonlocal nonlinear medium analogous to gravitational forces were reported
recently in~\cite{zeng}. In general, the soliton trajectories can be quite
complex as it is shown in Fig.~\ref{fig:dynamics}~(c), where a soliton is
launched with $g_x=g_y=5.89$. Since the soliton experiences an acceleration, it
radiates energy thus decreasing the effective power of the soliton and changing
the long-term dynamics~\cite{alberucci3}. Note that disregarding the long-term
effects of wave radiation that are naturally generated due to the acceleration
of the beam~\cite{Alberucci2016}, or the soliton breathing induced by the
corresponding approximation used, the predicted analytical results are
corroborated by full numerical propagation. Finally, we set the boundary
condition to be $N_b=900+50 \cos(\theta)$ with the same initial conditions
as~(c); the corresponding dynamics are shown in Fig.~\ref{fig:dynamics}~(d). In
this case, note that the equilibrium region is no longer centered around
$r=0$, nor it is of circular shape.

\begin{figure}[!htb]
\centerline{ \includegraphics[width=3.3in]{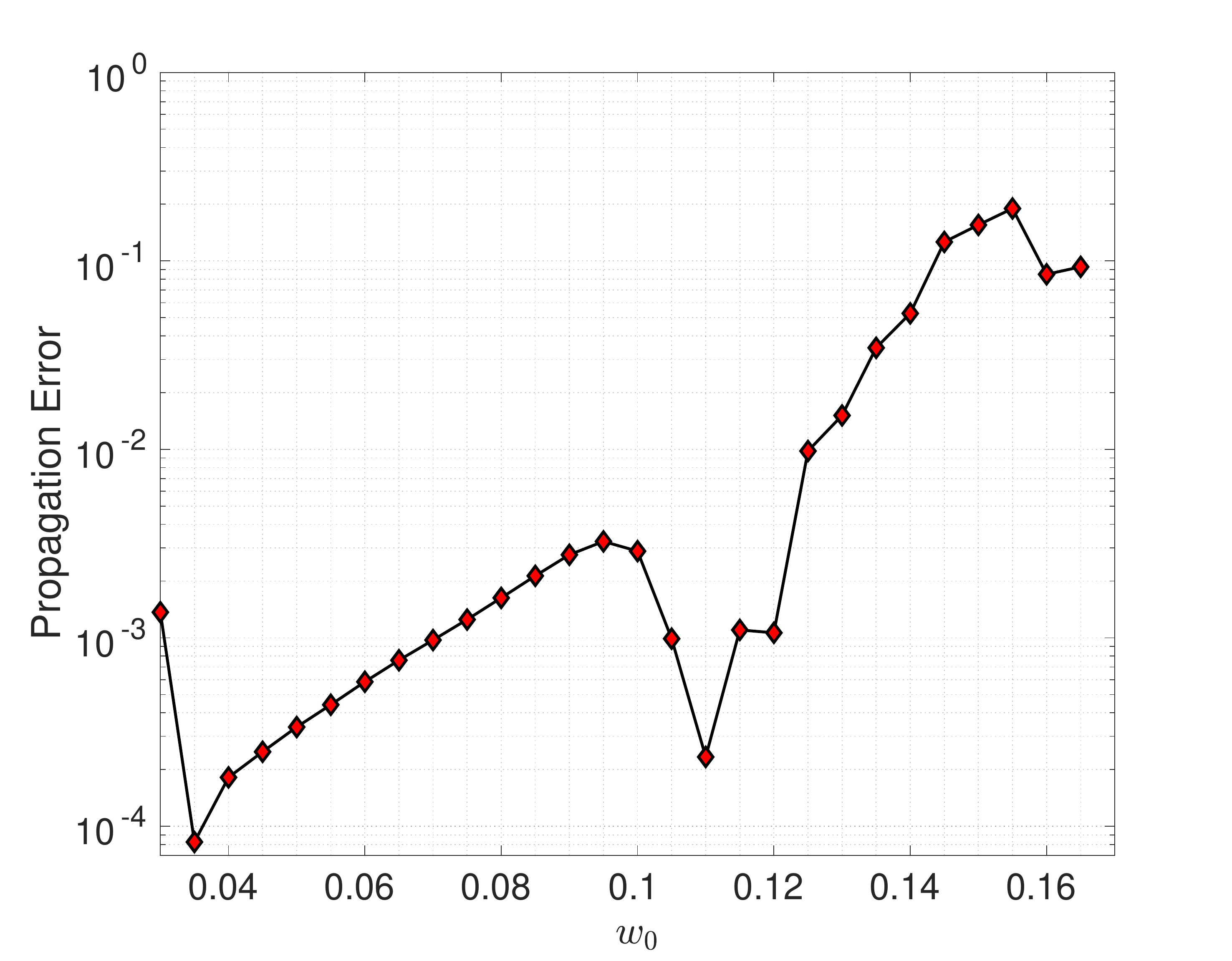} } 
\caption{Error in the soliton propagation as a function of the
    soliton's width. The error is measured as the distance between the position
    at a final time of the soliton propagated by the analytical model and the
    point of the soliton's centroid at that same time given by the numerical
    propagation. The propagation parameters, other than $\omega_0$, are the same
    as in Fig~\ref{fig:dynamics}~(d). Note that for $\omega_0\le 0.12$ we get
    errors of less than $10^{-2}$.}
\label{fig:prop_err}
\end{figure}

In order to validate the predicted trajectories as a function of the soliton
waist $\omega_0$ in a quantitative way, Fig.~\ref{fig:prop_err} shows the error
between the predicted final position of the soliton, and the one obtained
through the full numerical propagation (for the same propagation case as
in~Fig.~\ref{fig:dynamics}~(d)). For the particular case of $\rho=1$, the
approximation is better than $10^{-2}$ even for a beam waist of $0.12$ ($12\%$
of the radius of the boundaries).

A clear advantage of propagating
using~\eqref{eqn:main_result} over a full numerical propagation method, is that
with the analytical expressions, the computational time can be reduced by two or
even three orders of magnitude, without loosing significant accuracy. Through
comparison with a full numerical propagation of fundamental solitons with
relatively small waist (see~Fig.~\ref{fig:dynamics}), we found that this
simplified model describes very well the trajectories of the solitons, even for
relatively low power beams of about $P\sim700$ or higher, with the added benefit
of being able to draw conclusions about the dynamics from the analytical
expressions in~\eqref{eqn:main_result}.

\begin{figure}[!htb]
\centerline{
\includegraphics[width=3.3in]{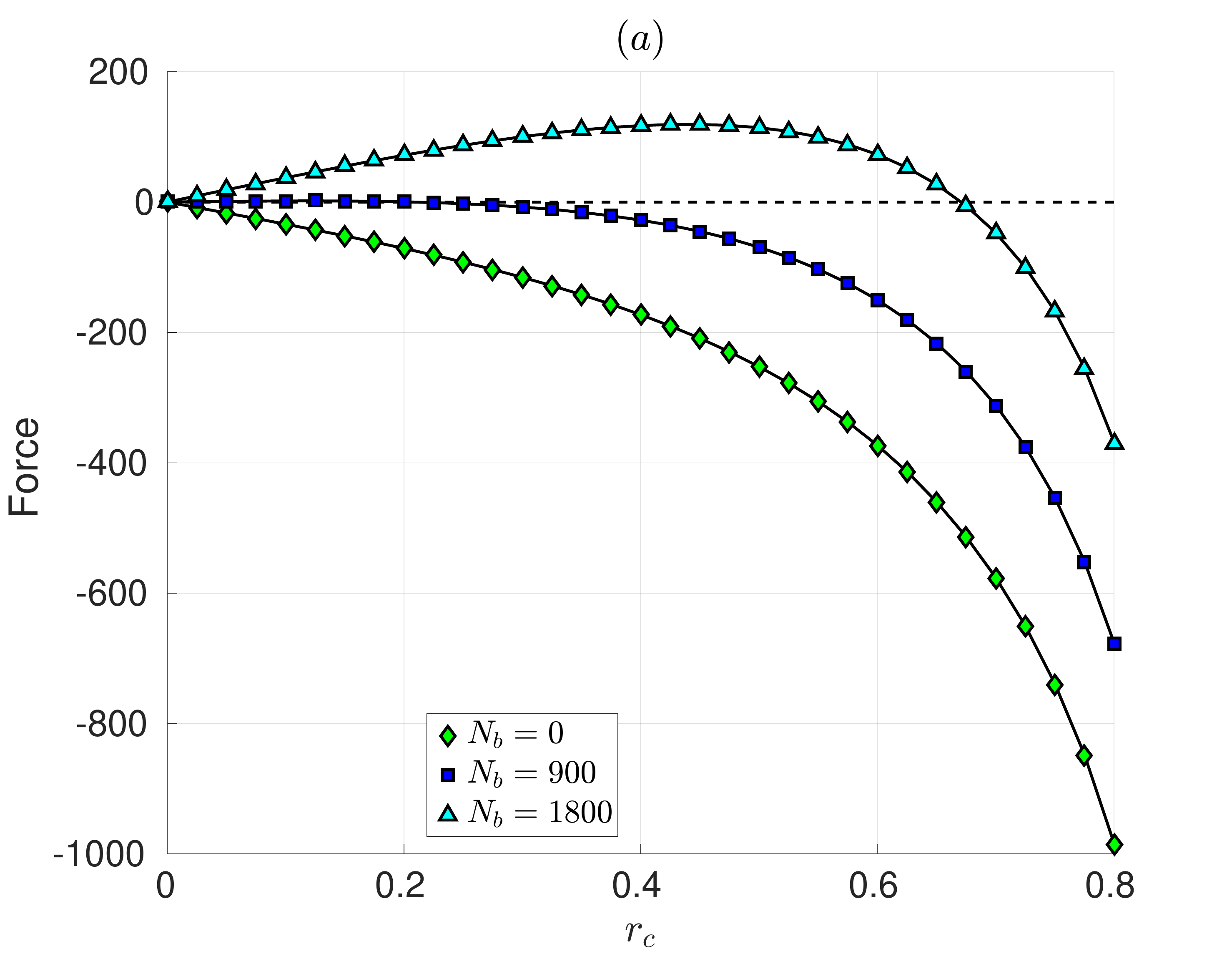} 
\includegraphics[width=3.3in]{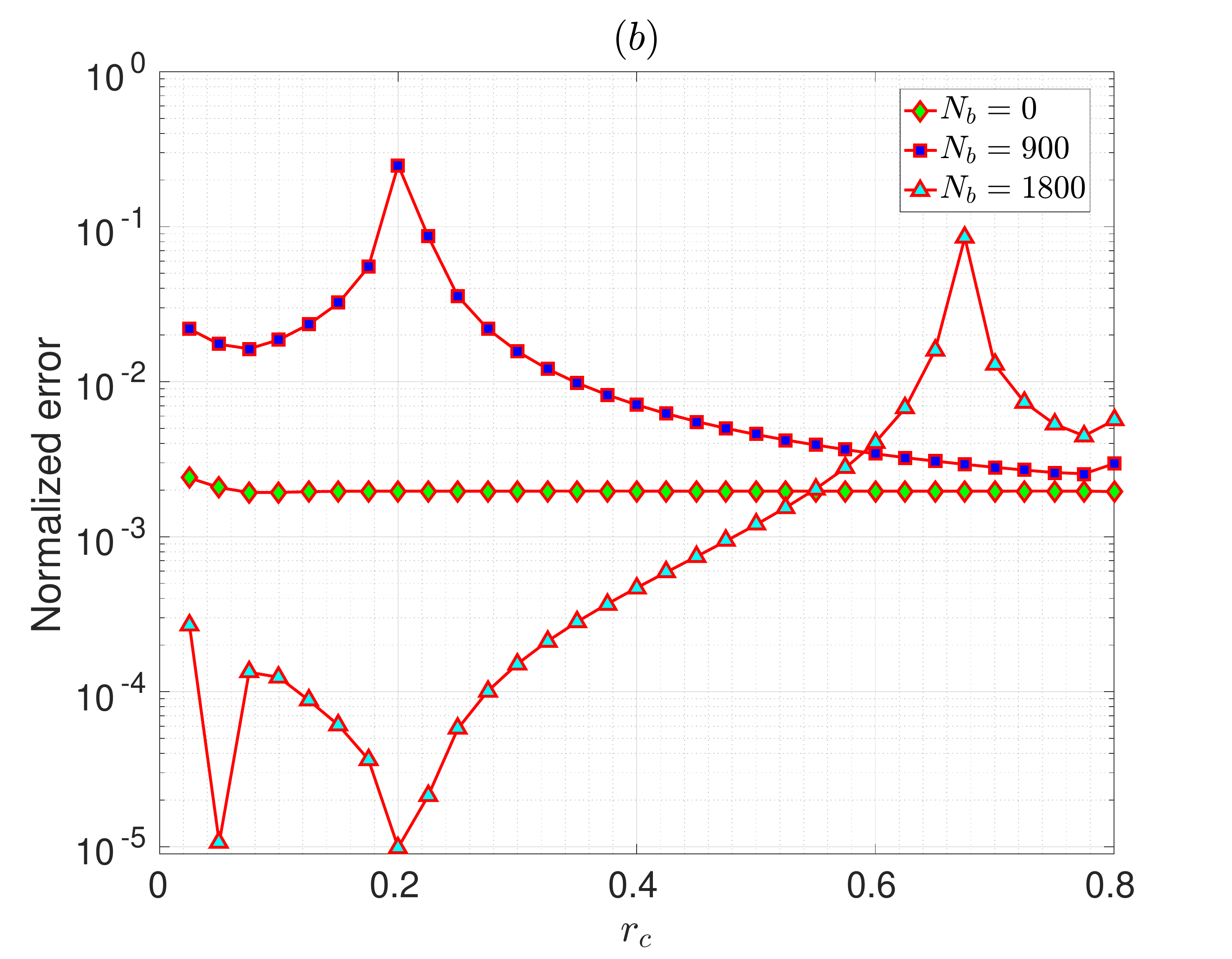} }
\caption{Comparison of the numerically computed force using~\eqref{eqn:forcenum}
  and the analytical approximation~\eqref{eqn:main_result} for a soliton with
  $P=3000$ and different values of $N_b$. In (a) the markers represent the
  values obtained by the numerical simulations, the black lines are the
  predicted values by the analytical model, wand the dashed black line is the
  force $=0$ axis. In (b) we present the normalized errors between the numerical
  and analytical forces, showing good agreement between simulation and the
  analytical prediction. Note that the peaks at around $r_c\approx 0.2$ and
  $r_c\approx 0.7$ are caused by the normalization value which is close is close
  to zero. }
\label{fig:compare}
\end{figure}

The quality of the approximation depends on how good the approximate force
formula in~\eqref{eqn:main_result} models~\eqref{eqn:forcenum}, thus
Fig.~\ref{fig:compare} shows a comparison between these two quantities for
different values of (constant) boundary conditions $N_b$, along with the
corresponding pointwise normalized error. It can be seen that the analytical
approximation yields a good approximation throughout the interior of the
circular cylinder.

\begin{figure}[!htb]
\centerline{
\includegraphics[width=3.3in]{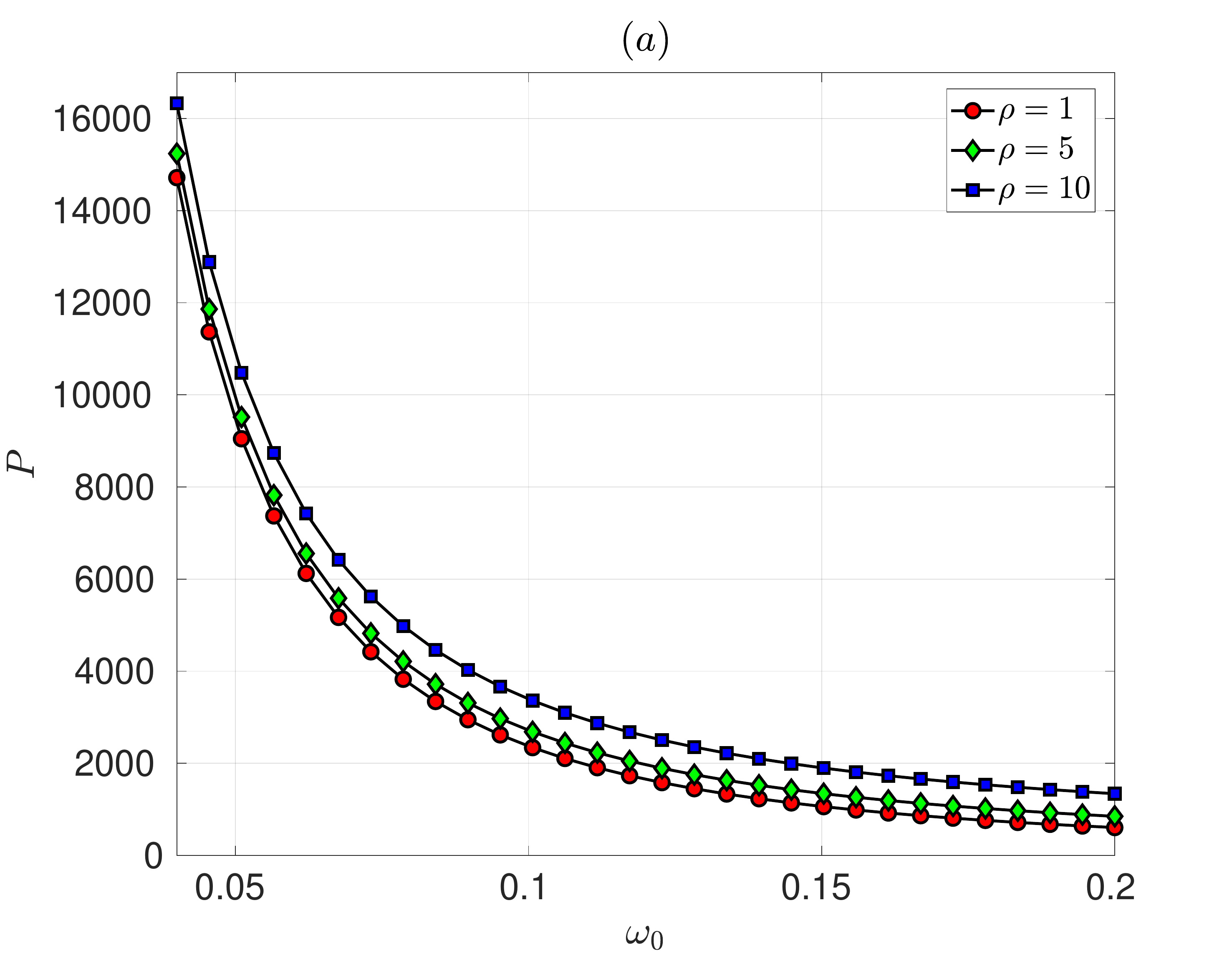} 
\includegraphics[width=3.3in]{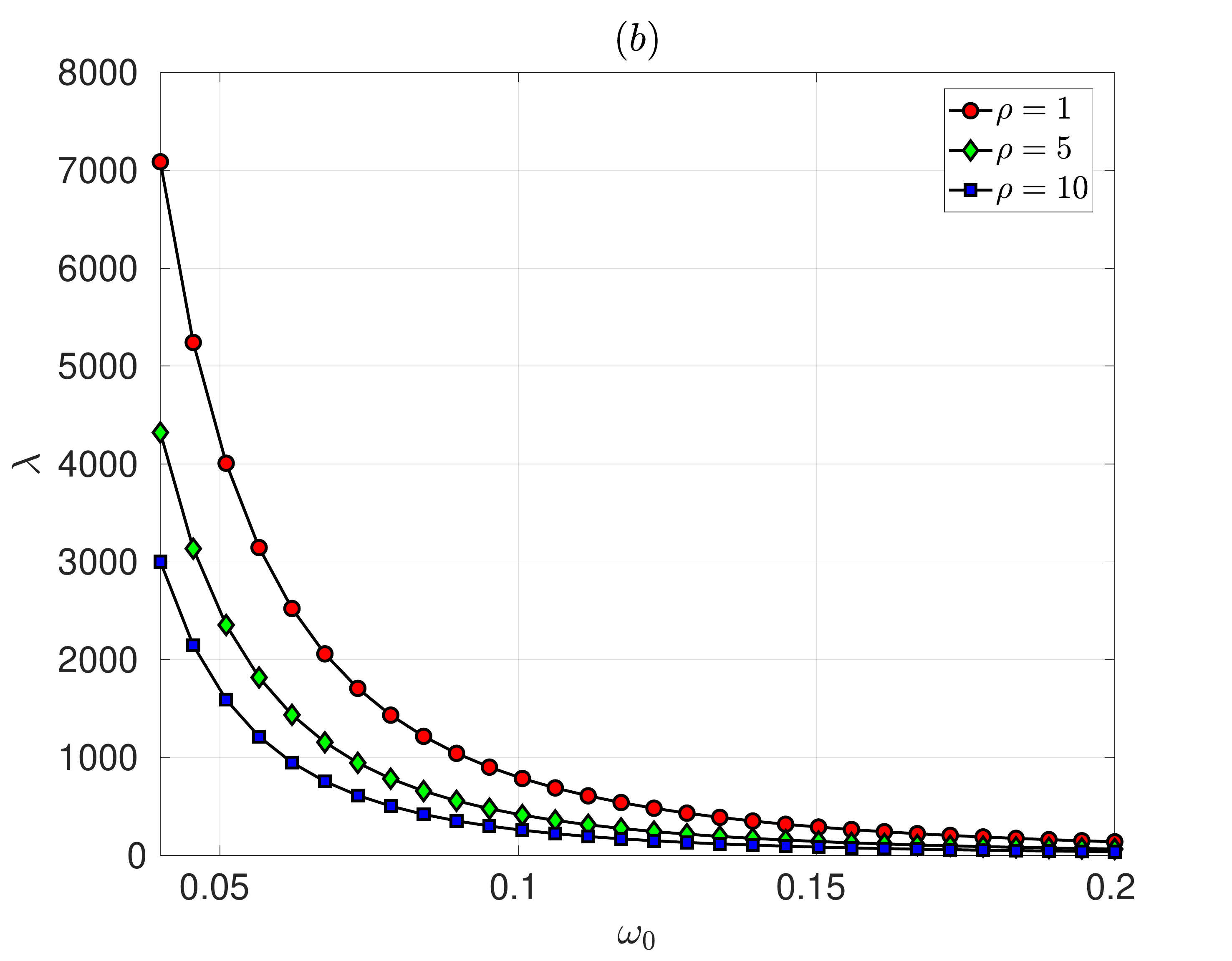} }
\caption{Soliton power (a) and propagation constant (b) as a function of
    the beam width. }
\label{fig:w0}
\end{figure}

The main soliton parameter in~\eqref{eqn:main_result} is the power, however the
validity of the approximation depends directly on the soliton waist. In order to
relate these two crucial parameters, Fig.~\ref{fig:w0}~(a) presents the soliton
power as a function of $\omega_0$ for relatively small waist solitons
($\omega_0 \le 0.2$). Fig.~\ref{fig:w0}~(b), on the other hand, presents the
propagation constant $\lambda$ also as a function of $\omega_0$.

\begin{figure}[!htb]
\centerline{\includegraphics[width=3.6in]{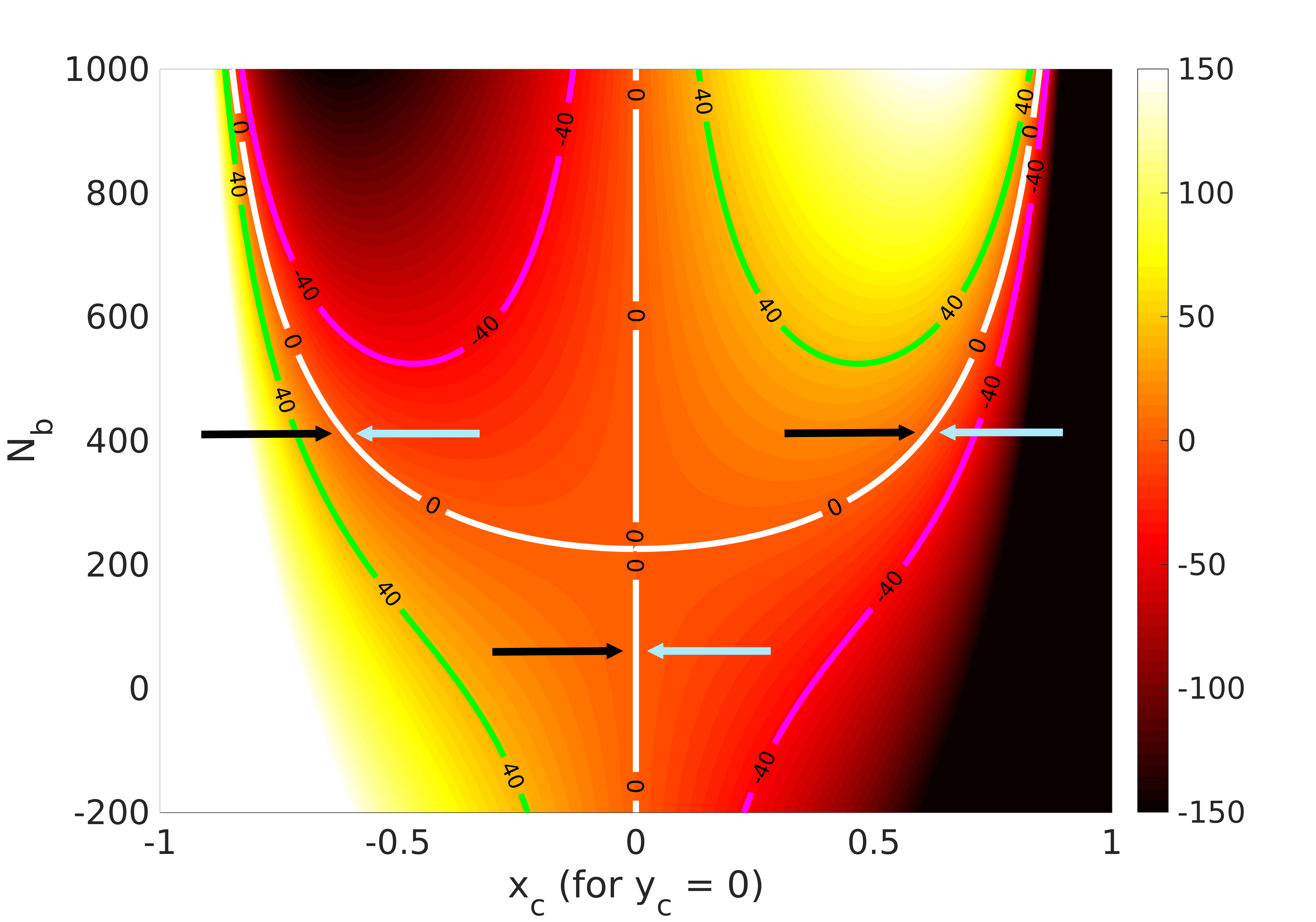}}
\caption{Supercritical pitchfork bifurcation as the boundary condition $N_b$
  (constant along the boundary) is increased. In this case, $P=800$ in a medium
  with $\rho=1$. The colormap represents the boundary force, and contour lines
  for force values of $0$ and $\pm 40$ are also present. After the bifurcation,
  the center equilibrium point becomes unstable, while the two appearing
  equilibrium points are stable.} \label{fig:bifur}
\end{figure}

\begin{figure}[!htb]
\centerline{ \includegraphics[width=3.3in]{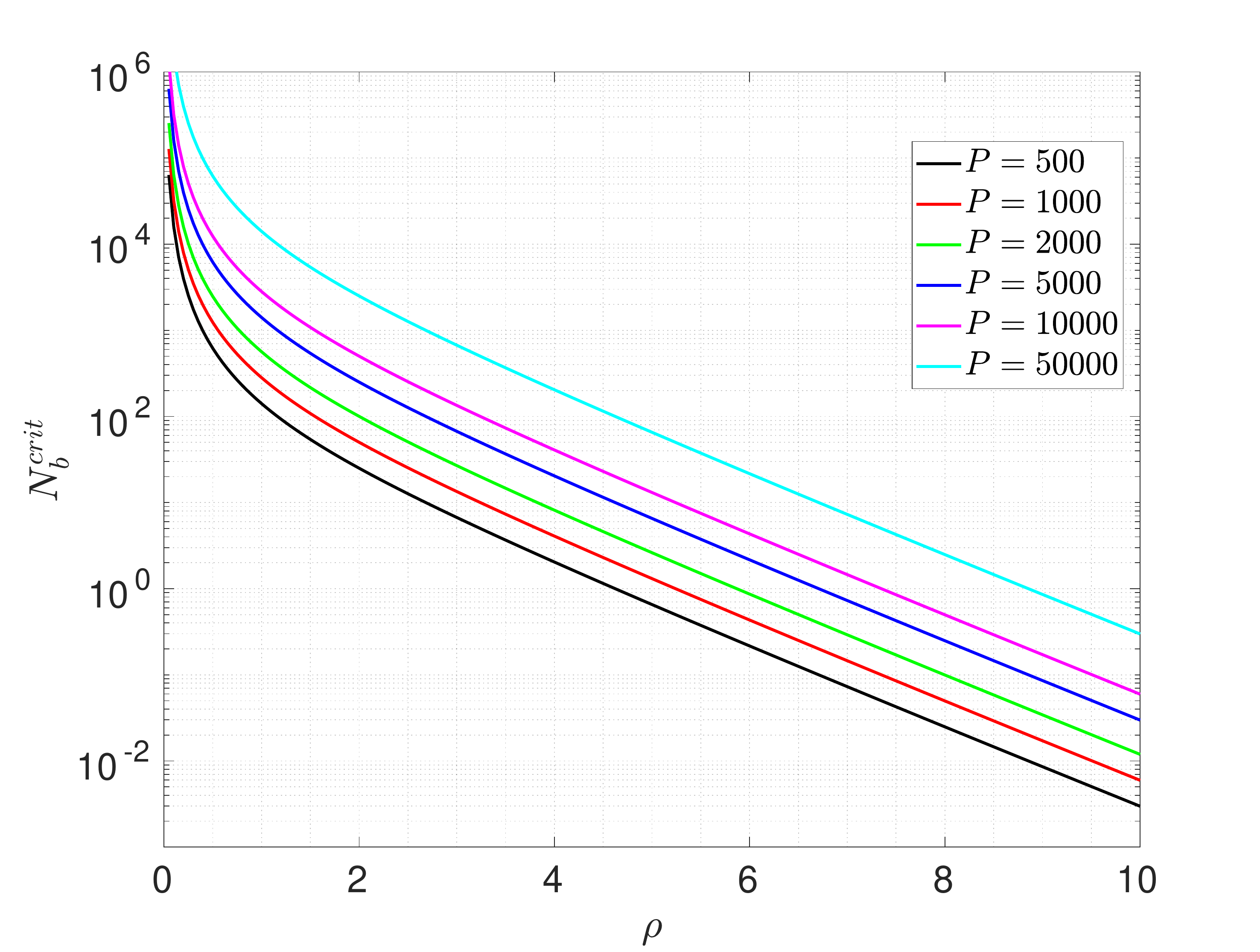} } 
\caption{Critical values of the boundary condition
    $N_b^{crit}$ as a function of the nonlocality parameter $\rho$ (semilog
    scale). For small $\rho$, the critical value of the boundary condition is
    large, and in fact, as $\rho \rightarrow 0$ we have
    $N_b^{crit}\rightarrow \infty$, as expected from the purely local case, for
    which no bifurcation is present. From~\eqref{eqn:main_result2}
    and~\eqref{eqn:main_result2} we see that for a fixed $\rho$, the dependence
    on $P$ of $N_b^{crit}$ is linear.}
\label{fig:rho_vs_nb}
\end{figure}

The examples in Fig.~\ref{fig:dynamics} and Fig.~\ref{fig:compare} show that
when a non-zero boundary condition is present, the dynamical system that
describes the soliton trajectory experiences a bifurcation of the center
equilibrium point. In particular, a supercritical pitchfork
bifurcation~\cite{perko1991differential} is present when a constant boundary
condition is applied, that is, as we increase the value of the boundary
condition, the stable equilibrium point at the center degenerates into two
equilibrium points, with the center point being unstable. In
Fig.~\ref{fig:bifur} we show the bifurcation diagram for this case, where
instead of using $\nrc$, we took $y_c=0$ and plotted the diagram with respect to
$x_c$ in order to illustrate the pitchfork behavior. The critical value of the
boundary condition $N_b^{crit}$ at which the bifurcation occurs can be obtained
as the value of $N_b$ for which the derivative of the force with respect to
$N_b$ becomes zero at $x_c=0$ (with $y_c=0$), or equivalently by finding the
value of $N_b$ that gives a zero eigenvalue on the linearization matrix of the
dynamical system~\cite{perko1991differential}. In Fig.~\ref{fig:rho_vs_nb} we
present the behavior of $N_b^{crit}$ as a function of $\rho$, which shows that
for small values of $\rho$, i.e. the highly nonlocal case, $N_b^{crit}$ tends to
be high, while for $\rho>1$ the behavior is exponentially decaying. This result
is consistent with the highly nonlocal case~\cite{Shou2009a} for which the only
equilibrium point is at the center of the cylinder, independently of the
boundary condition. Additionally, Fig.~\ref{fig:rho_vs_nb} shows the
behavior for different values of the soliton's power. Note that for constant
$\rho$, the dependence of $N_b^{crit}$ as a function of $P$ is linear as a
result from~\eqref{eqn:main_result2} and~\eqref{eqn:main_result2}. For
angle-dependent boundary conditions, we observed a similar behavior for which a
new stable equilibrium region appear after a threshold value is exceeded (see
Fig.~\ref{fig:dynamics}~(d) right column).

Finally, we show that in the limiting case $\rho\rightarrow 0$, and for constant
boundary conditions, we recover the same result as the one reported
in~\cite{Shou2009a}. Using the asymptotic forms of the modified Bessel functions
for small arguments~\cite{abramowitz}, and taking the limit as $\rho$ goes to
zero we get
\begin{align}
  \displaystyle \mathbf{F}_2(r_c) = -\frac{P}{2\pi r_c} \sum_{n=1}^\infty
  r_c^{2n}\; \mathbf{\hat r} = -\frac{P}{2\pi} \frac{r_c}{1-r_c^2} 
  \; \mathbf{\hat r}, \quad \text{for } \rho\rightarrow 0,
\end{align}
which is exactly Eq.~(8) from~\cite{Shou2009a} (for constant boundary
conditions, $\mathbf{F}_H \rightarrow 0$ as $\rho \rightarrow 0$). In this
particular case, the force is independent of the (constant) boundary condition,
and only under an angle-dependent boundary condition does a force in the angular
direction appears, which is given by 
\begin{align}
  \displaystyle \mathbf{F}_H(r_c,\theta_c) = \sum_{n=-\infty}^\infty 
  i \: n \:  r_c^{n-1} \: A_n e^{i n \theta_c} \;
  \boldsymbol{ \hat{ \theta } }, \quad \text{for } \rho\rightarrow 0.
\end{align}
This shows that the bifurcation of the equilibrium point does not occur in the
highly nonlocal case, demonstrating that the introduction of the parameter
$\rho$ in the model yields significantly different soliton dynamics.

\section{Conclusions}
In summary, we have obtained an analytical approximation for the force exerted
on a soliton propagating in a finite nonlocal nonlinear media with a circular
cylinder shape, with an arbitrary degree of nonlocality. This approximation can
be used to predict the trajectories of small-waist soliton beams
($\omega_0 << 1$) and it is in good agreement with the numerical experiments
presented. When the magnitude of the boundary condition exceeds a certain
critical value, the original equilibrium point at the origin degenerates into an
unstable equilibrium point and a stable equilibrium curve where the force is
equal to zero; in the case of constant boundary conditions, this bifurcation is
described by a supercritical pitchfork bifurcation. Recently, some analogies
between optical soliton interactions and the dynamics of galactic cores in the
scalar field dark-matter scenario have been reported
\cite{navarrete2017spatial}. We hope that a similar treatment to the one
reported here can shed some light in this direction.

\section*{Funding} Consejo Nacional de Ciencia y Tecnolog\'{i}a (CONACYT) (243284).

\pagebreak
\bibliographystyle{abbrv}
\bibliography{finite_nonlocal}

\end{document}